 \documentclass[preprint,3p,times,twocolumn]{elsarticle}

\usepackage{amssymb}

\journal{Physics Letters B}

\begin{document}

\begin{frontmatter}



\title{PMT Test Facility at MPIK Heidelberg and Double Chooz Super Vertical Slice}


\author{J. Haser}
\author{F. Kaether\corref{cor1}}
\cortext[cor1]{Corresponding authors. Tel.: +49 6221 516829\\
{\it Email:} kaether@mpi-hd.mpg.de, reinhold@mpi-hd.mpg.de}
\author{C. Langbrandtner}
\author{M. Lindner}
\author{B. Reinhold\corref{cor1}}
\author{S. Sch\"{o}nert}
\address{Max-Planck-Institut f\"{u}r Kernphysik, Saupfercheckweg 1,
D-69117 Heidelberg, Germany}

\begin{abstract}
Proceedings supplement for conference poster at Neutrino 2010, Athens, Greece. The poster is available at {\tt http://www.mpi-hd.mpg.de/personalhomes/kaether/Neutrino2010\_kaetherreinhold.pdf}
\end{abstract}


\end{frontmatter}

\begin{figure}[htb]
\begin{center}
\includegraphics[width=1.03\columnwidth]{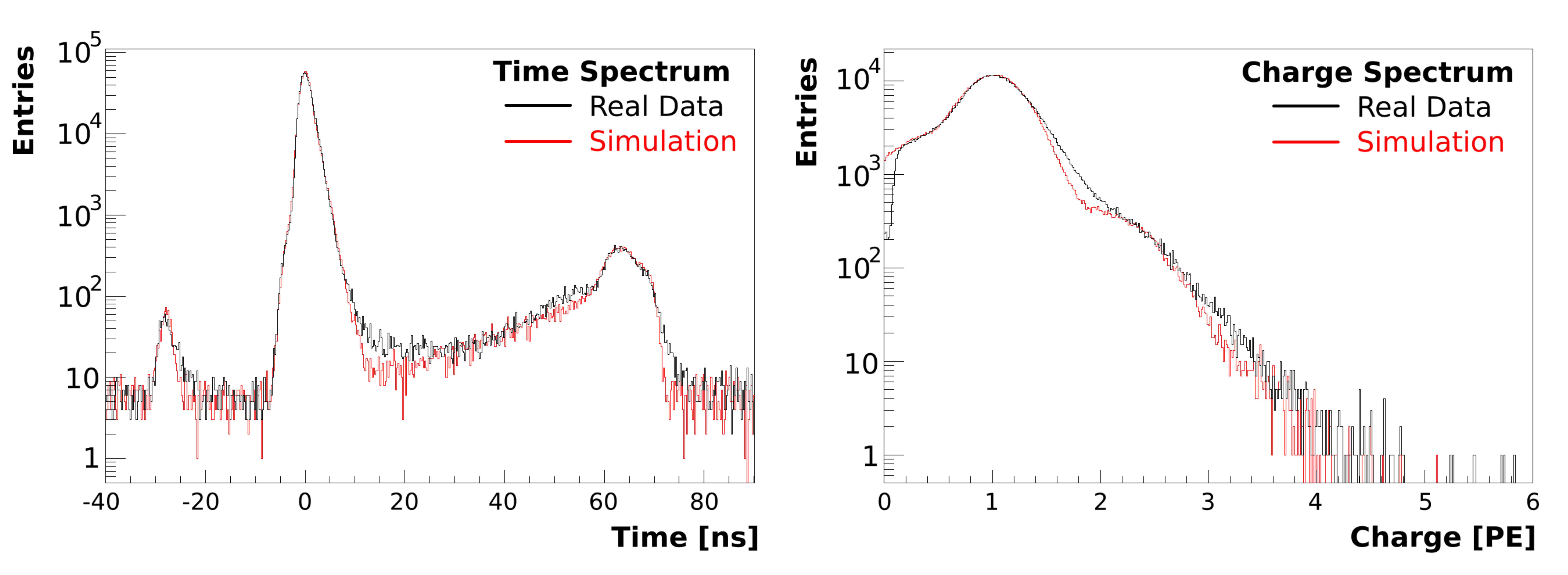}
\caption{Time spectrum (left) and charge spectrum (right) of one typical inner detector PMT (measured and simulated data).}
\label{fig:1}
\end{center}
\end{figure}

The two inner detectors of Double Chooz (DC) \cite{DC} will both be observed by 390 photomultiplier tubes (PMTs) to record light pulses produced by neutrino induced reactions inside the scintillator volumes. The task of calibrating the PMTs was shared among DC Japan and MPIK/ RWTH Aachen. At Heidelberg a \textit{test facility} was built which allows to calibrate 30 PMTs simultaneously. The VME based data acquisition (DAQ) includes devices for charge spectrum recording, for time measurements and scalers for dark rate counting. The PMT calibration campaign encompassed the determination of the single photo electron (SPE) gain and resolution, transit time spread and linearity for multi-PE events.
Two trigger boards developed by RWTH Aachen were integrated to test their performance and to examine afterpulse probabilities and dark rate counting. The derived results were translated into time and charge probability density functions (see fig.\ \ref{fig:1}), which in turn were implemented into the Double Chooz Monte Carlo simulation.
 
The \textit{Super Vertical Slice (SVS)} is an upgrade of the PMT test facility. It consists of the
complete and original Double Chooz electronics plus an acrylic vessel of 30 liters filled with the
Gadolinium doped $\nu$-Target scintillator. The electronics consists of PMTs, high voltage (HV) power supply, HV splitter, frontend electronics (FEE), trigger system (TS), waveform digitizers (WFD) and DAQ.
It has been successfully used to finalize the interface of FEE and TS, such as signal gains, and to control the quality of the final FEE production.
The electronics response in case of very high energy depositions has been studied. The obtained results hint at the feasibility of using muon-induced isotopes, such as $^{12}$B or Michel electrons, for detector calibration.
The SVS has been first to record pulses from the original $\nu$-Target scintillator with the original Double Chooz electronics thereby demonstrating its functionality in a realistic multichannel environment before the final installation at Chooz.

\section*{Acknowledgements}
For providing hardware and support during the PMT calibration campaign we
thank S.\ Lucht and A.\ St\"{u}ken (RWTH Aachen) and DC Japan.

We thank all Double Chooz collaboration members who provided generous
support for the SVS by contributing hardware and their wisdom:
TS (RWTH Aachen), FEE (Drexel University, US), HV power supply (Tohoku University), HV splitters (CIEMAT
Madrid), WFDs (APC Paris), acrylic vessel (CEA Saclay), scintillator (MPIK
Heidelberg).
J. Dawson and A. Cabrera (APC Paris) provided DAQ software and critical
support for the SVS. Finally all efforts would have been impossible without the
mechanical and electrical workshops at MPIK Heidelberg.



\end{document}